\documentstyle[12pt]{article}
\newcommand{\be}{\begin{equation}}
\newcommand{\ee}{\end{equation}}
\newcommand{\ba}{\begin{eqnarray}}
\newcommand{\ea}{\end{eqnarray}}
\newcommand{\bb}{\begin{thebibliography}}
\newcommand{\eb}{\end{thebibliography}}
\newcommand{\ci}[1]{\cite{#1}}
\newcommand{\bi}[1]{\bibitem{#1}}
\newcommand{\lab}[1]{\label{#1}}

\begin{document}

\vspace{2cm}
\begin{flushright} {\large JINR Rap. Comm. 2-94}
\end{flushright}
\vspace{1cm}

\begin{center}
{ \large The $A_{LL}$ asymmetry in diffractive  reactions
           and  structure of quark-pomeron vertex } \\
\vspace{1cm}

S.V.Goloskokov, O.V.Selyugin\\
Bogolubov Laboratory of Theoretical Physics,\\
Joint Institute for Nuclear Research, Dubna\\
%Head Post Office P.O.Box 79, 101000 Moscow, Russia \\
%----------------------the footnote-----------------------------------
\begin{figure}[b]
\begin{center}
\rule{175mm}{.25mm}
\small
%Tel.:  (7)(096)(21) 65\ 696 ---
%Telex: 911621 DUBNA SU ---
%Fax: (7)(096)(21) 65\ 084 \\
E-mail:  goloskkv@thsun1.jinr.dubna.su; selugin@thsun1.jinr.dubna.su
%{\footnotesize \em and}
%selugin@main1.jinr.dubna.su
\end{center}
\end{figure}
\end{center}

\vspace{1cm}

\begin{abstract}
 Theoretical predictions for the behavior of $A_{LL}$ asymmetry
determined by the pomeron-hadron vertex are done. Strong
dependence of the asymmetry on the mass creating quarks and
transfer momenta is shown.
\end{abstract}

\section{Introduction}

The spin effects in high energy reactions are still
an open problem of perturbative QCD. Like in the framework of
the standard perturbative QCD, there is no way to explain
many  spin phenomena revealed in different processes at high energies.

    The spin-flip effects do not exist  in the massless limit
\ci{brod} in hard processes where all quark masses must be omitted.
If we do not neglect the quark masses, the spin-flip amplitudes are
suppressed  as a power of s with respect to the spin-non-flip ones.

          The  $t$-channel exchange with vacuum quantum numbers (pomeron)
gives a fundamental contribution to high energy reactions
at fixed momenta transfer ($s \to \infty,\; t$-fixed).
The calculations of diagrams and their summations are usually
performed in the leading logarithmic approximation (see e.g \ci{lip}).
However, the spin-flip amplitudes are absent in this approximation.

As has been intimated, the most part of spin experimental data
at high energies are obtained at fixed momenta transfer.
So, the theoretical research of the pomeron spin structure is very important.
   The vacuum $t$-channel amplitude is usually associated in QCD with the
two-gluon exchange \ci{low}. The properties of the spinless pomeron were
analysed in \ci{la-na,don-la} on the
basis of a QCD model with taking account of nonperturbative properties of
the theory.

A similar model was used to investigate the spin effects
in pomeron exchange.  It was demonstrated that different contributions
determined by a gluon ladder \ci{gol1} and quark loops \ci{gol2}
may lead to the spin-flip amplitude growing as s in the limit $s \to \infty$.
Factorization of the $qq$ amplitude was shown
into the spin-dependent large-distance part and the high-energy spinless
pomeron. This permits us to define the quark-pomeron vertex and to discuss
the results of summation of the pomeron ladder graphs in higher orders of QCD.
It has been shown that the obtained amplitude leads to the ratio of spin-flip
and  non-flip amplitudes being perhaps independent of the energy \ci{gol-pl}.
This result should modify different spin
asymmetries and lead to new effects in high energy diffractive reactions
which can be measured in future experiments in the RHIC at Brookhaven
\ci{bunce}.

\section{Pomeron and spin phenomena}

 It was demonstrated in the soft momentum transfer region in \ci{gol1}
within the qualitative
QCD analysis that the $qq$ spin-flip
amplitude growing as $s$ can be obtained  in the $\alpha_s^3$ order.
Factorization of the spin-flip amplitude has been shown into
the spin-dependent large-distance part (quark-pomeron vertex) and the
high-energy spinless pomeron.
The  quantitative calculations of the spin effects in $qq$ scattering were
performed in the $\alpha_s^3$ order in the half-hard region $s \to \infty,
\; |t|>1GeV^2$ \ci{gol4} where the perturbative theory can be used.

It was shown that the quark-pomeron vertex (Fig.1) in the perturbative region
has a form
\be
V_{qqP}^{\mu}(k,q)=\gamma_{\mu} u_0(q)+2 m k_{\mu} u_1(q) +2 k_{\mu}
/ \hspace{-2.3mm} k u_2(q) + i \frac{u_3(q)}{2}\epsilon^{\mu\alpha\beta\rho}
k_\alpha q_\beta \gamma_\rho \gamma_5+im\frac{u_4(q)}{2}   \lab{ver}
\sigma^{\mu\alpha} q_\beta.
\ee
In (\ref{ver}) $u_i(q)$ are the vertex functions. The term
proportional to $\gamma_{\mu}$  corresponds to the standard spinless
pomeron that reflects the well-known fact that the spinless quark-pomeron
coupling is like a $C=+1$ isoscalar photon  \ci{la-na}. We use the
simple form of the $u_0(q)$ vertex function
$$   u_0(q)=\frac{\mu_0^2}{\mu_0^2+Q^2} ,\;\;\;q^2=-Q^2,$$
with $\mu_0 \sim 1Gev$ introduced in  \ci{don-la}. The functions
$u_1(q) \div u_4(q)$ at large $Q^2$ were calculated in perturbative QCD
\ci{gol4}. Their magnitudes are not very small.
Additional spin-flip contributions to the quark pomeron
vertex can be connected with instanton effects (see \ci{do},\ci{fo} e.g).
The magnitude of these effects is not very well defined because they are
model dependent.

 Note that the structure of the quark-pomeron vertex function (\ref{ver}) is
drastically different from the  standard spinless pomeron. Really, the terms
$u_1(q)-u_4(q)$ lead to the spin-flip in the quark-pomeron vertex in contrast
to the term proportional to $u_0(q)$ which is spin-non-flip one. As a result
the new terms can modify different spin asymmetries and lead to new
effects in high energy reactions.

To show this, let us investigate the quark pair productions in
diffractive hadron reactions  This sort of reactions was investigated
by different authors (see \ci{berg} e.g.). We shall estimate the longitudinal
double spin asymmetries in the reaction (Fig.2) as an example. Note that to
extract this process, we must detect the final proton with the
longitudinal momenta $p^{'}_2 = x_f p_2 $. The angle of this final proton
in the c.m.s. is determined by the relation
$ \sin(\theta_{p^{'}_2})=\sqrt{Q^2/(x_f s)} $
and for $\sqrt{s}=100GeV, Q^2=10GeV, x_f=.7$ we have $\sin(\theta_{p^{'}_2})
\sim 0.04$.

The resulting asymmetry looks as follows
\be
A_{ll}=C_g \frac{(1-x_f^2) \{4 u_0^2+Q^2 u_3 [2 u_0+4 m^2 u_1+2 m^2u_2 +m^2 u_4
]\}}{(1+x_f^2) \ln(s(1-x_f)/Q^2) [4 u_0^2+u_3^2 \; Q^4/2]}.        \lab{asy}
\ee
Here
\be
 C_g= \frac{\int_{0}^{1} \Delta g(y) dy}{(y g(y))|_{y=0}}, \lab{c}
\ee
$m$ is the quark mass.
For the simple form of the gluon structure function
$$  g(y)=\frac{3}{y} (1-y)^5          $$
we have for the coefficient (\ref{c})
\be
 C_g= \frac{\int_{0}^{1} \Delta g(y) dy}{3}. \lab{i3}
\ee
It is well known that
$\int_{0}^{1} \Delta g(y) dy \sim 3 $
is necessary for the explanation of different spin effects \ci{sof}.
In this case $C_g \sim 1$.

\section{The $A_{LL}$ asymmetry  for the energy of RHIC and AGS}

It is easy to see from (\ref{asy}) that for the standard spinless pomeron
that contains the $u_0$ term, only the asymmetry (\ref{asy}) has a slow $Q^2$
dependence on the logarithmic term in the denominator. However, we must
observe a strong $Q^2$ dependence in the $A_{ll}$ asymmetry for the pomeron
vertex with the spin-flip part. The resulting asymmetry is equal to zero
for $x_f=1$.
The $A_{ll}$ asymmetry for energy of RHIC $\sqrt s = 100 GeV$
estimated on the basis of
perturbative results for vertex functions
for $x_{F}=.7$ and $C_{g}=1$
is shown in Fig.3. The
resulting asymmetry can reach $10 \div 12 \% $ in the case of large magnitude
for the integral (\ref{i3}).  As can be seen, from Fig. 3,  we have
an insignificant negative
asymmetry for light guarks and positive for quarks with large mass.
However, for low energies the asymmetry of light quarks can be
sufficiently large.

    Similar experiments can be performed at AGS energies ($\sqrt s =7 GeV$).
In this case we can analyse the production of light and charm quarks.
The predictions for the pomeron contribution to the $A_{LL}$ asymmetries
in this case
for $x_{F}=.7$ and $C_{g}=1$
are shown in Fig. 4. The asymmetry is large and reaches
 40\% because in this case we have no large suppression by
the $\ln[ s(1-x)/Q^{2}]$ term in (\ref{asy}).
 We can see that asymmetry of light quarks can reach $15 \%$.
The maximum asymmetry strongly depends on scattering energies, $Q^{2}$
and quark mass.
So, for $\sqrt s = 100 GeV$ we have the maximum asymmetry at
$Q^{2} = 20 \div 25 GeV^{2}$ for the quark mass equal to $1.3 GeV$
and at $Q^{2} = 45 \div 50 GeV^{2}$ for the quark mass equal to $4.6 GeV$

\section{Conclusion}

So, the $A_{LL}$ asymmetry can be measured and the information about
the spin structure of the quark-pomeron vertex can be extracted.
Therefore, the experiment has been carried out to determine the dependence
of asymmetry $A_{LL}$ on $Q^{2}$ for separate mass quark,
for example, $m_{q} = 1.3 GeV$, or the dependence of asymmetry on
the quark mass at separate $Q^{2}$, for example, for
$\sqrt s = 100 GeV$ $Q^2 = 20 \div 25 GeV^2$
This asymmetry can be used for the evaluation in RHIC of
$$\int_{0}^{1} \Delta g(y) dy$$
if the magnitudes of vertex
functions $u_1(q) \div u_4(q)$ are known from other experiments or in the
opposite  case (when the $\Delta g(y)$ is known) for the determination of the
vertex functions $u_1(q) \div u_4(q)$.
This permits one to determine a relative magnitude of the
nonperturbative instanton contribution.

Moreover, using the handedness method \ci{efr} the final quark-spin
correlations with the spin of initial hadrons can be observed.

\vspace{.2cm}

     {\it Acknowledgement.}  {\hspace {0.5cm}}   The authors express
  their deep gratitude
  to S.P.Kuleshov and A.N. Sissakian
  for support of this work and to A.V. Efremov, O.V.Teryaev, M.V.Tokarev and
  Yu.A. Panebratzev  for fruitful discussions.

%\newpage

\vspace{3cm}
%\newpage
{\large {Figure Captions}}
\vspace{2cm}
\phantom{.}\\
Fig. 1   Quark - pomeron vertex.\\
\\
Fig. 2   Diagram for the production of a quark pair
         in pomeron-hadron interaction.\\
\\
Fig. 3   $A_{ll}$ - asymmetry in the case of the pomeron-hadron
	 interaction at the RHIC \\
\phantom{.} \hspace{1cm}         energies. \\
\phantom{.} \hspace{1cm}   (for the spinless vertex - solid curve;\\
\phantom{.} \hspace{1cm} for the spin-flip vertex  and different mass of
creating quarks:\\
\phantom{.} \hspace{1cm} dotted line - m = $0.005 GeV$,
                              dot-dached line - $m = 1.3 GeV$, \\
\phantom{.} \hspace{1cm} short dashed line - $ m = 4.6 GeV$).\\
\\
Fig. 4   $A_{ll}$ - asymmetry in the case of the pomeron-hadron
	 interaction at the AGS \\
\phantom{.} \hspace{1cm}          energies. \\
\phantom{.} \hspace{1cm}   (for the spinless vertex - solid curve;\\
\phantom{.} \hspace{1cm} for the spin-flip vertex  and different mass of
creating quarks:\\
\phantom{.} \hspace{1cm} dotted line - m = 0.005 GeV,
                              dot-dached line - $m = 1.3 GeV$, \\

\end{document}